
     \magnification=\magstep1

     \def\bar{\overline}
     
     \def\dsl{\raise .15ex\hbox{/}\kern-.57em\partial}
     \def \Dsl{\,\raise .15ex\hbox{/}\mkern-13.5mu D}\hfill\break
\centerline{{\bf Ising Model Scaling Functions at Short Distance}}
\vskip.1in\par\noindent
\centerline{\sl John Palmer}
\centerline{\sl Department of Mathematics }
\centerline{\sl University of Arizona }
\centerline{\sl Tucson, AZ 85721}
\centerline{\sl palmer@math.arizona.edu}
\vskip.1in\par\noindent
I will sketch a proof that the short distance behavior
of the even scaling functions for the Ising model that
arise by taking the scaling limit of the Ising correlation
functions from below the critical temperature is given
by the Luther--Peschel formula [6] (see below).  The fact
that the Luther--Peschel formula is consistent with 
conformal field theory insights into the correlations
for the large scale limit of the critical Ising model has
lead to the conviction in the Physics community that
this formula is right [4]. The {\it critical scaling limit\/} for 
the Ising model is 
obtained by sitting right at $T_c$ (the critical temperature) 
and extracting the large
scale limit of the correlations at this fixed temperature. 
However, as far as I am aware
there is no mathematical proof that the large scale
limit of the critical Ising model even exists (with the 
exception of the asymptotics of the diagonal two point
correlation at the critical point [15]).  In this
note we will not address the question of the critical
(massless) scaling limit for the Ising model but instead we will
consider the massive scaling limit obtained by looking
at the lattice correlations at the scale of the correlation
length as the temperature {\it approaches\/} the critical 
temperature.  This scaling limit was first considered in 
[14]
and [11] and a mathematical proof of its existence was 
established in [7].  What we will sketch a proof of here 
is that the short distance asymptotics of this scaling 
limit for
the even scaling function from below $T_c$ is given by the 
Luther--Peschel formula.  There is a 
conjecture in the Physics literature called the scaling
hypothesis which suggests that the large scale asymptotics
at the critical temperature should agree with the short
distance behaviour of the massive scaling theory.  A 
rather detailed version of this scaling hypothesis has 
been proved for the two point function [12],[13].  Modulo
the fact that this scaling hypothesis can only be 
literally correct for the scaling limit from above $T_c$ 
(since the critical Ising correlations vanish for an
odd number of spins on the lattice and the odd 
scaled correlations from below $T_c$ are non-zero) this
hypothesis does suggest a connection between the result
we consider here and the conjectured result for the
critical scaling limit.  Incidentally, the technique to be explained 
below should extend to an analysis of the short distance 
asymptotics of the odd correlations in the $T<T_c$ scaling
limit.  

In [11] M.~Sato, T.~Miwa and M.~Jimbo showed that the
scaled correlations for the Ising model have logarithmic 
derivatives that can be expressed in terms of the
solution to a monodromy preserving deformation 
problem. In one version of this connection the log
derivative is expressed in terms of the solution to
non-linear monodromy preserving deformation equations.  
Of more technical interest for
us is an alternative connection with the {\it linear\/} differential
equation which gives rise to the monodromy preserving
deformation.  Much recent work on the asymptotics
of non-linear isomonodromic equations exploits in a 
precisely analogous manner the connection with an 
associated linear problem (see [3] and references for 
an account of modern Riemann--Hilbert techniques in
the analysis of such problems).

I will begin by recalling the Sato, Miwa, 
Jimbo characterization [11] for the logarithmic derivative of 
the scaled correlations for the two dimensional Ising 
model (a more detailed proof of this result appears in [8]).  
The two 
dimensional Ising model on the lattice ${\bf Z}^2$ is a classical
statistical mechanical 
system of $\pm 1$ spins with ferromagnetic nearest neighbor
interactions.  The Boltzmann weight for a configuration
of spins $\sigma_i$ at sites $i\in\Lambda$ with $\Lambda$ a finite subset of $
{\bf Z}^2$ is
given by 
$$e^{K\sum_{<ij>}\sigma_i\sigma_j},$$
where the sum is over nearest neighbors $<ij>$ in $\Lambda$ 
($K>0$ favors nearest neighbor alignment and is 
referred to as the ferromagnetic case).
To be a little more precise if the sum is restricted to
nearest neighbors
$<ij>$ both of which are in $\Lambda$ we will refer to {\it open\/} boundary 
conditions.  If the nearest neighbors $<ij>$ are allowed
one element (say $i$) on the boundary of $\Lambda$ (but not in $\Lambda$) 
and we extend the spin configuration to the boundary by
setting 
$\sigma_i=+1$ then we will refer to {\it plus\/} boundary 
conditions. 
Write
$$<\sigma_{i_1}\sigma_{i_2}\cdots\sigma_{i_n}>_T,\eqno (1.1)$$
for the thermodynamic (infinite volume) limit of the 
spin correlation functions at temperature $T>0$.  For
$T$ less than the critical temperature $T_c$ we consider the
limit of the finite volume correlations with {\it plus }
boundary conditions and for $T>T_c$ we consider the limit
with {\it open\/} boundary conditions (these are the limits 
discussed in [8]).  Next we introduce a scaling parameter
$\delta >0$ and a temperature $T(\delta )<T_c$ so that the correlation
length (the exponential rate of clustering for (1.1)) is $\delta^{
-1}$ 
and we define for $a_j\in {\bf R}^2,$
$$\tau_{-}(a_1,a_2,\ldots ,a_n)=\lim_{\delta\rightarrow 0}{{<\sigma_{
\delta^{-1}a_1}\sigma_{\delta^{-1}a_2}\cdots\sigma_{\delta^{-1}a_
n}>_{T(\delta )}}\over {<\sigma >_{T(\delta )}^n}},\eqno (1.2)$$
where $<\sigma >_T=<\sigma_i>_T>0$ is the translation invariant one
point function (the spontaneous magnetization).  The 
existence of this limit and the 
subtleties associated with the fact that $\delta^{-1}a_j$ is not
always a lattice point in ${\bf Z}^2$ are discussed in [7]. We will
refer to $\tau_{-}$ as the scaling function for the Ising model
from below $T_c$.  We define a scaling function, $\tau_{+},$ for the
Ising model from above $T_c$ by
$$\tau_{+}(a_1,a_2,\ldots ,a_n)=\lim_{\delta\rightarrow 0}{{<\sigma_{
\delta^{-1}a_1}\sigma_{\delta^{-1}a_2}\cdots\sigma_{\delta^{-1}a_
n}>_{T(\delta )^{*}}}\over {<\sigma >_{T(\delta )}^n}},\eqno (1.3
)$$
where $T(\delta )^{*}$ is the Kramers-Wannier dual temperature
$T(\delta )^{*}>T_c$ (see [7]).  Note that in the denominator it is 
$T(\delta )<T_c$ 
which appears and not the dual temperature; this is 
significant 
since $<\sigma >_T=0$ for $T>T_c.$ 

In [11] Sato, Miwa and Jimbo introduced a characterization 
of $d_a\log\tau_{\pm}(a)$ which we will now describe for $\tau_{-}$.  For later
developments it will be useful to introduce a parameter
$m>0$  which will scale the dependence on the $a_j$ in (1.2)
and (1.3). For Sato, Miwa and Jimbo the functions $\tau_{\pm}$ are
{\it tau functions\/} for  monodromy preserving 
deformations of the Euclidean Dirac equation on ${\bf R}^2$.  The
appropriate Dirac operator is,
$$mI-\dsl=\left[\matrix{m&-2\partial\cr
-2\bar{\partial}&m\cr}
\right],$$
where
$$\eqalign{\partial ={1\over 2}\left({{\partial}\over {\partial x_
1}}-i{{\partial}\over {\partial x_2}}\right),\cr
\bar{\partial }={1\over 2}\left({{\partial}\over {\partial x_1}}+
i{{\partial}\over {\partial x_2}}\right).\cr}
$$
Although we will {\it not\/} be working exclusively with 
holomorphic functions, the presence of $\partial$ and $\bar{\partial}$ in
the Dirac operator makes it very convenient to identify
${\bf R}^2$ with ${\bf C}$ in the usual way, $(x_1,x_2)\rightarrow 
z=x_1+ix_2$ and of 
course,
$\bar {z}=x_1-ix_2$.  Henceforth we will use this identification
and work over {\bf C} instead of ${\bf R}^2$.  For brevity we will
write $f(z)$ for a function of the two real variables $x_1$ and 
$x_2$, even though it is customary to use a notation like
 $f(z,\bar {z})$ to avoid the temptation to regard $f(z)$ as a 
holomorphic function of $z$.  

Let ${\bf a}=\{a_1,a_2,\ldots ,a_n\}$ denote a collection of $n$ {\it distinct }
points in ${\bf C}$.  Let ${\cal R}({\bf C}\backslash {\bf a})$ denote the 2 fold ramified
covering space of ${\bf C}\backslash {\bf a}$ with base point $\alpha_
0$ covering 
$a_0\in {\bf C}$ with $a_0\ne a_j$ for $j=1,2,\ldots n$.  Let $\gamma_
j$ denote
a simple closed loop based at $a_0$ surrounding $a_j$ but
no other point $a_k$ for $k\ne j$.  Let $\Gamma_j$ denote the deck
transformation of ${\cal R}({\bf C}\backslash {\bf a})$ associated with $
\gamma_j$.

The differential operator $mI-\dsl$ acts naturally
on smooth functions $f:{\cal R}({\bf C}\backslash {\bf a})\rightarrow 
{\bf C}^2$.  We will call a 
smooth function $w:{\cal R}({\bf C}\backslash {\bf a})\rightarrow 
{\bf C}^2$ an {\it Ising type\/} solution
to the Dirac equation if and only if
$$\left(mI-\dsl\right)w=0,$$
and $w\circ\Gamma_j=-w$ for $j=1,\ldots n$ (that is $w$ changes by a sign ``going 
around''  the loop $\gamma_j$). An Ising type solution to the 
Dirac equation has local expansions (essentially Fourier
expansions) about each of the points $a_{\nu}$ given by
$$w(z)=\sum_{k\in {\bf Z}}c_k^{\nu}(w)w_k(z_{\nu})+c_k^{\nu}{}^{*}
(w)w^{*}_k(z_{\nu}),$$
where $z_{\nu}=z-a_{\nu}$, the quantities $c^{\nu}_k(w)$ and $c^{
\nu *}_k(w)$ are
complex numbers which we refer to as {\it local expansion}
{\it coefficients,\/} and the functions $w_k(z)$ and $w_k^{*}(z)$ are 
multivalued solutions
to the Dirac equation defined by (see [11]),
$$w_k(z)=\left[\matrix{v_k(z)\cr
v_{k+1}(z)\cr}
\right]=\left[\matrix{{{m^{k-{1\over 2}}z^{k-{1\over 2}}}\over {2^{
k-{1\over 2}}\Gamma (k+{1\over 2})}}\cr
{{m^{k+{1\over 2}}z^{k+{1\over 2}}}\over {2^{k+{1\over 2}}\Gamma 
(k+{3\over 2})}}\cr}
\right]+O(m^{k+{1\over 2}}),\eqno (1.4)$$
$$w_k^{*}(z)=\left[\matrix{\bar {v}_{k+1}(z)\cr
\bar {v}_k(z)\cr}
\right]=\left[\matrix{{{m^{k+{1\over 2}}\bar {z}^{k+{1\over 2}}}\over {
2^{k+{1\over 2}}\Gamma (k+{3\over 2})}}\cr
{{m^{k-{1\over 2}}\bar {z}^{k-{1\over 2}}}\over {2^{k-{1\over 2}}
\Gamma (k+{1\over 2})}}\cr}
\right]+O(m^{k+{1\over 2}}).\eqno (1.5)$$
Where $v_k(z)=e^{i(k-{1\over 2})\theta}I_{k-{1\over 2}}(mr)$,
$z=re^{i\theta}$ is the polar form for $z$, $I_s$ is the modified
Bessel function of order $s$, and $\Gamma$ is the usual gamma 
function.  The functions $w_k(z)$ and $w_k^{*}(z)$ are 
simultaneously solutions to the Dirac equation 
$(mI-\dsl)w_k^{(*)}=0$ and eigenfunctions for the infinitesimal
rotation that commutes with the Dirac operator. 
Observe that the functions $w_k(z)$ and $w_k^{*}(z)$ become
less and less locally singular as functions of $z$ as the
index $k$ increases.  Whenever we write down a local
expansion that explicitly exhibits only a few terms
it is understood that the remaining terms $+\cdots$ involve
less locally singular wave functions $w_k^{(*)}$ (i.e., higher values of the 
index $k$) 
than the terms which make an explicit appearance. 

If we
define $C=\left[\matrix{0&1\cr
1&0\cr}
\right]$ then it is easy to see that $w_k^{*}$ is 
obtained 
from $w_k$ by the following conjugation,
$$w_k^{*}(z)=C\overline {w_k(z)}.$$
$$ $$
The conjugation defined on $f:{\bf C}\rightarrow {\bf C}^2$ by $f
\rightarrow f^{*}:=C\bar {f}$ will have 
special significance for us as it commutes with the 
Dirac operator and defines a {\it real\/} structure on the space 
of solutions to the Dirac equation.  This real structure 
plays a role in the Ising model application we are looking
at; observe that because the monodromy 
multiplier -1 is {\it real\/} the conjugation $*$ also defines a real
structure on the space of Ising type solutions to the 
Dirac equation. 

Now we recall some results from [11].  Write 
$x\cdot y=x_1y_1+x_2y_2$ for the standard bilinear form on
$ $${\bf C}^2$, so that the standard Hermitian form is given by
$\bar {x}\cdot y.$ If $u$ and $v$ are Ising 
type solutions to the Dirac equation then the inner 
product $\bar {u}\cdot v$ descends to a function on ${\bf C}\backslash 
{\bf a}$ since the
monodromy multipliers cancel out.  We will
say that such a solution, $w,$ is in $L^2({\bf C})$ if and only if
$${i\over 2}\int_{{\bf C}}\bar {w}\cdot w\,dzd\bar {z}<\infty .$$
It is a result of SMJ [11] the space of Ising type solutions
to the Dirac equation that are in $L^2({\bf C})$ is an $n$
dimensional vector space.  The local expansions for such 
a solution, $w,$ are restricted,
$$w(z)=\sum_{k=0}^{\infty}c_k^{\nu}(w)w_k(z_{\nu})+c_k^{\nu *}(w)
w^{*}_k(z_{\nu})\hbox{\rm \ for }\nu =1,\ldots ,n,$$
and there is a formula for the inner product of two 
such $L^2$ solutions in terms of the leading local expansion
coefficients,
$${i\over 2}\int_{{\bf C}}\bar {u}\cdot v\,\,dzd\bar {z}=-{4\over {
m^2}}\sum_{\nu}\overline {c_0^{\nu}(u)}c_0^{\nu *}(v)=-{4\over {m^
2}}\sum_{\nu}\overline {c_0^{\nu *}(u)}c_0^{\nu}(v),\eqno (1.6)$$
which follows from the identities,
$${i\over 2}\bar {u}\cdot v\,dzd\bar {z}=-{i\over m}d(\bar {u}_2v_
1dz)={i\over m}d(\bar {u}_1v_2d\bar {z}),\eqno (1.7)$$
that are a consequence of $u$ and $v$ satisfying the Dirac
equation, and the use of Stokes' theorem to localize
the integrals of the differentials on the right of (1.7) to the 
singularities $a_{\nu}.$  Equation (1.6) shows that $L^2$ Ising type
solutions are determined uniquely by the low order 
expansion coefficients $c_0^{\nu}$, $\nu =1,\ldots n$ (or the coefficients
$c_0^{\nu *},$$\nu =1,\ldots ,n$).  We will now characterize the $
L^2$ Ising 
type solution we are interested in by specifying the
behavior of its low order expansion coefficients.  This 
solution will satisfy a ``reality'' condition with respect
to the conjugation $*$ so it will be convenient to 
work with a real basis constructed from $w_k$ and $w_k^{*}$.
Define $\Re w_k={1\over 2}(w_k+w_k^{*})$ and 
$\Im w_k={1\over {2i}}(w_k-w_k^{*}).$ Then for each $\nu =1,\ldots 
n$ there exists
a unique Ising type solution ${\cal W}_{\nu}$ to the Dirac equation
which is real $({\cal W}_{\nu}^{*}={\cal W}_{\nu})$, is in $L^2$,
and which has leading order local expansions,
$${\cal W}_{\nu}(z)=\delta_{\mu\nu}\Im w_0(z_{\mu})+T_{\mu\nu}\Re 
w_0(z_{\mu})+\cdots ,\eqno (1.8)$$
for $z$ near $a_{\mu}$.
The coefficient matrix $T_{\mu\nu}$ is real since ${\cal W}_{\nu}$ is real but
as we shall see it is not necessary to specify anything
further about it to achieve uniqueness (it will turn out to
be skew symmetric). 
  We will consider the 
existence and uniqueness for ${\cal W}_{\nu}$ in a moment but first
we note that the connection with the scaling function $\tau_{-}$ for the Ising
model is,
$$d_a\log\tau_{-}(ma)={m\over {2i}}\sum_{\nu}c_1^{\nu}({\cal W}_{
\nu})da_{\nu}-\overline {c_1^{\nu}({\cal W}_{\nu})}d\bar {a}_{\nu}
.\eqno (1.9)$$
This result is due to SMJ [11] (see also the proof of 
theorem 5.1 in [8]) and will be the foundation of the analysis we
make of the $m\rightarrow 0$ limit of $\tau_{-}(ma)$ that concerns us in 
this paper.  Note that in [11] it is $i{\cal W}_{\nu}$ that is taken as the
fundamental wave function and this alters the formula
(1.9) just a little.

We now recall the construction of the wave
function ${\cal W}_{\nu}$ and the proof that it is uniquely 
determined by the conditions given above. In [11], SMJ 
prove that there exists a canonical basis $\{{\bf w}_{\nu}\}$ for the space 
of $L^2$ Ising
type solutions to the Dirac equation.  This basis is 
characterized by local expansions with leading order 
terms,
$${\bf w}_{\nu}(z)=\delta_{\mu\nu}w_0(z_{\mu})+c_0^{\mu *}({\bf w}_{
\nu})w_0^{*}(z_{\mu})+\cdots ,\eqno (1.10)$$
for $z$ near $a_{\mu}.$  Applying the conjugation $*$ to (1.10) we get
what is called by SMJ the dual canonical basis, 
$${\bf w}_{\nu}^{*}(z)=\delta_{\mu\nu}w_0^{*}(z_{\mu})+\overline {
c_0^{\mu *}({\bf w}_{\nu})}w_0(z_{\mu})+\cdots\eqno (1.11)$$
The formula for the inner product (1.6) shows that $L^2$ 
Ising type solutions are uniquely determined by the
values of the $n$ linear functionals $c_0^{\nu}$ for $\nu =1,\ldots 
,n$.  
Hence we find,
$${\bf w}_{\nu}(z)=\sum_{\alpha}c_0^{\alpha *}({\bf w}_{\nu}){\bf w}_{
\alpha}^{*}(z),\eqno (1.12)$$
and
$$\sum_{\alpha}c_0^{\alpha *}({\bf w}_{\nu})\overline {c_0^{\mu *}
({\bf w}_{\alpha})}=\delta_{\mu\nu}.\eqno (1.13)$$
If we use the formula (1.6) to calculate the inner product
of two elments of the canonical basis {\bf w}e find,
$$<{\bf w}_{\mu},{\bf w}_{\nu}>={i\over 2}\int_{{\bf C}}\bar {{\bf w}}_{
\mu}\cdot {\bf w}_{\nu}\,dzd\bar {z}=-{4\over {m^2}}c_0^{\mu *}({\bf w}_{
\nu})=-{4\over {m^2}}\overline {c_0^{\nu *}({\bf w}_{\mu})}\eqno 
(1.14)$$
Now write $c_0^{*}$ for the matrix with $\mu\nu$ matrix element
$c_0^{\mu *}({\bf w}_{\nu})$ and we can summarize (1.13) and (1.14) by saying
that $c_0^{*}$ is an hermitian symmetric, negative definite 
matrix with $\overline {c_0^{*}}c_0^{*}=I$.  If we now rewrite the local 
expansion result for ${\cal W}_{\nu}$ in terms of $w_0$ and $w_0^{
*}$  instead
of $\Im w_0$ and $\Re w_0$ we find by comparing local expansions
that,
$$\eqalign{{\cal W}_{\nu}={1\over 2}\sum_{\mu}(T_{\mu\nu}-i\delta_{
\mu\nu}){\bf w}_{\mu},\cr
{\cal W}_{\nu}={1\over 2}\sum_{\mu}(T_{\mu\nu}+i\delta_{\mu\nu}){\bf w}_{
\mu}^{*}.\cr}
\eqno (1.15)$$
Comparing these two results with (1.12) we find,
$$c_0^{*}(T-iI)=(T+iI),\eqno (1.16)$$
where $T$ denotes the matrix with $\mu\nu$ matrix element $T_{\mu
\nu}.$
We can solve this for $T$ to get
$$T=i(c_0^{*}-I)^{-1}(c_0^{*}+I).\eqno (1.17)$$
The fact that $c_0^{*}$ is negative definite implies that $c_0^{*}
-I$ 
is invertible so we can {\it define} $T$ by (1.17).  Since $c_0^{
*}$ is
Hermitian symmetric it follows that $T^{*}=-T$. Since
$\overline {c_0^{*}}c_0^{*}=I$ it follows that $\bar {T}=T$, that is $
T$ is real. Thus
$T$ must be (real) skew symmetric as well.  With this 
definition for $T$ we can {\it define} ${\cal W}_{\nu}$ by (1.15) and 
uniqueness follows from the developments leading up to
(1.17). 

We are interested in computing the 
$m\rightarrow 0$ limit of the right hand side of (1.9). In order to do
this we will first introduce a slightly different 
characterization of ${\cal W}_{\nu}$ which will introduce the principal
player in our understanding of the $m\rightarrow 0$ limit; this is
an appropriate Green function for the Dirac operator
$mI-\dsl$ acting on a suitable space of ``multivalued''
functions (or more precisely it is an explicit formula for
the $m=0$ limit of this Green function).  We begin by 
noting a uniqueness result 
which will suggest the definition of the Green function
we are interested in.  Suppose that $w$ is an $L^2$ Ising type
solution to the Dirac equation which has
leading order local expansions,
$$w(z)=c^{\mu}(w)\Re w_0(z_{\mu})+\cdots ,\eqno (1.18)$$
for $z$ near $a_{\mu}$.  Then $w$ is identically 0. To see this 
observe that for such a $w$ one has 
$c_0^{\mu}(w)=c_0^{\mu *}(w)={{c^{\mu}(w)}\over 2}.$ Substituting this result in the
formula for $<w,w>$ obtained from (1.6) one finds that
$<w,w>$ must be negative unless $c^{\mu}(w)=0$ in which case
$<w,w>=0$.  

Roughly speaking our strategy in dealing
with ${\cal W}_{\nu}$ can now be described as follows.  We will make
a {\it local\/} subtraction $\phi_{\nu}$ from ${\cal W}_{\nu}$ which will kill off the
term $\delta_{\mu\nu}\Im w_0$ in the expansion (1.8) for ${\cal W}_{
\nu}.$ The 
difference ${\cal W}_{\nu}-\phi_{\nu}$ then satisfies the inhomogenous Dirac
equation,
$$(mI-\dsl)({\cal W}_{\nu}-\phi_{\nu})=-(mI-\dsl)\phi_{\nu}:=f_{\nu}
,\eqno (1.19)$$
and ${\cal W}_{\nu}-\phi_{\nu}$ is uniquely characterized by this equation and 
the fact that it is in $L^2({\bf C})$ with local expansions of type 
(1.18) at each of the points $a_{\mu}$.  We will demonstrate the 
convergence of the solution to (1.19) in the limit 
$m\rightarrow 0$ with sufficient control to say what happens to (1.9)
in this limit. 

We will now sketch the
existence theory needed to make this possible (we follow
the existence theory in SMJ [11] quite closely).  Let 
$C_0^{\infty}({\cal R}({\bf C}\backslash {\bf a}))$ denote the space of complex valued $
C^{\infty}$ functions, $f$, on
${\cal R}({\bf C}\backslash {\bf a})$ such that $f\circ\Gamma_j=-
f$, and such that the projection
of the support of $f$ in ${\cal R}({\bf C}\backslash {\bf a})$ onto $
{\bf C}\backslash a$ is compact.  Let
${\cal H}^{1,0}_m$ denote the Hilbert space completion of 
$C_0^{\infty}({\cal R}({\bf C}\backslash {\bf a}))$ with respect to the norm determined by the
inner product,
$$<f,g>_{1,0}={i\over 2}\int_{{\bf C}}\left(\overline {\partial f
(z)}\partial g(z)+m^2\overline {f(z)}g(z)\right)dzd\bar {z}.\eqno 
(1.20)$$
Let ${\cal H}_m^{0,1}$ denote the Hilbert space completion of 
$C^{\infty}_0({\cal R}({\bf C}\backslash {\bf a}))$ with respect to the inner product
$$<f,g>_{0,1}:={i\over 2}\int_{{\bf C}}\left(\overline {\bar{\partial }
f(z)}\bar\partial g(z)+m^2\overline {f(z)}g(z)\right)dzd\bar {z}.\eqno 
(1.21)$$
Now suppose that $f\in C_0^{\infty}({\cal R}({\bf C}\backslash {\bf a}
))$ and consider the linear
functional,
$$C_0^{\infty}({\cal R}({\bf C}\backslash {\bf a}))\ni g\rightarrow{
i\over 2}\int\overline {f(z)}g(z)dzd\bar {z}.$$
For $m\ne 0$ it is clear that this linear functional extends 
to a {\it continuous\/} linear functional on both ${\cal H}_m^{1,
0}$ and ${\cal H}^{0,1}_m$.  
Thus by the
Riesz representation theorem there exists an $F^{ij}\in {\cal H}_
m^{i,j}$  so 
that, 
$$<F^{i,j},g>_{i,j}={i\over 2}\int_{{\bf C}}\overline {f(z)}g(z)d
zd\bar {z},\eqno (1.22)$$
where $i,j=1,0$ or 0,1. 
It follows from this last equation that away from the
branch points $F^{i,j}$  is a distribution solution to
$$(m^2-\Delta )F^{i,j}=f.$$
Elliptic regularity implies that $F^{i,j}$ is $C^{\infty}$ ``away from the
branch points'' (i.e., on the covering space ${\cal R}({\bf C}\backslash 
{\bf a})$).
Near the branch point $z=a_{\nu}$, $F^{i,j}$ is an $L^2$ branched solution 
to the 
Helmholtz equation $(m^2-\Delta )F^{i,j}=0$ and so has a local
expansion,
$$F^{i,j}(z)=\sum_{k\ge 0}a_k^{\nu ,i,j}v_k(z-a_{\nu})+\sum_{k\ge 
0}b_k^{\nu ,i,j}\bar {v}_k(z-a_{\nu}),\eqno (1.22)$$
Since $\partial F^{1,0}$ is locally in $L^2$ it follows that $a_0^{
\nu ,1,0}=0$ and
since $\bar{\partial }F^{0,1}$ is locally in $L^2$ it follows that $
b_0^{\nu ,0,1}=0$ for
$\nu =1,\ldots ,n$.

Now we
apply this to solve the Dirac equation,
$$(m-\dsl)F=f,\eqno (1.23)$$
where $f=\left[\matrix{f_1\cr
f_2\cr}
\right]$ and $f_j\in C_0^{\infty}({\cal R}({\bf C}\backslash {\bf a}
))$.  Let $F^{0,1}\in {\cal H}^{0,1}_m$ be the
solution to $(m^2-\Delta )F^{0,1}=f_1$ constructed above and let
$F^{1,0}\in {\cal H}_m^{1,0}$ be the similar solution to $(m^2-\Delta 
)F^{1,0}=f_2$.
Define,
$$F:=\left[\matrix{mF^{0,1}+2\partial F^{1,0}\cr
2\bar{\partial }F^{0,1}+mF^{1,0}\cr}
\right].\eqno (1.24)$$
It is easy to check that $F$ is a solution to the Dirac 
equation (1.23) and furthermore that because $F^{i,j}$ lies
in ${\cal H}^{i,j}_m$ it follows that the solution $F$ is locally in $
L^2$. 
This is not quite the solution to (1.23) which we want.  
Instead, notice that because $F$ is locally
in $L^2$ about each point $a_{\nu}$ we can subtract a linear
combination of the canonical wave functions ${\bf w}_{\nu}$ from
$F$ to produce a new solution ${\bf F}$ to (1.23) with the 
property that,
$$c_0^{\nu}({\bf F})=0\hbox{\rm \ for }\nu =1,\ldots ,n\eqno (1.2
5)$$
Observe that these boundary conditions uniquely 
determine an $L^2$ solution to (1.23) since the formula
(1.6) shows that any $L^2$ solution to the homogenous
equation killed by all the linear functionals $c_0^{\nu}$ is
necessarily 0. Now that we know existence and 
uniqueness for $L^2$ solutions to (1.23) with the boundary
conditions (1.25) we can use this enlarge the class of
suitable
boundary conditions to include the conditions relevant
to the Ising model.  Let $N$ denote the space of $L^2$ Ising type
solutions, $w$, to $(m-\dsl)w=0$.  Let ${\cal N}$ denote the image
of $N$ under the map,
$$N\ni w\rightarrow (c_0(w),c_0^{*}(w))\in {\bf C}^n\oplus {\bf C}^
n,$$
where $c_0^{(*)}(w)$ denotes the $n$ tuple $(c_0^{1(*)}(w),\ldots 
,c_0^{n(*)}(w))$
in ${\bf C}^n$.
We know that ${\cal N}$ is an $n$ dimensional subspace of 
${\bf C}^n\oplus {\bf C}^n$
and that the intersection of ${\cal N}$ with $\{0\}\oplus {\bf C}^
n$ is $\{0\}$.
Since $\{0\}\oplus {\bf C}^n$ is $n$ dimensional it follows that $
{\cal N}$ and
$\{0\}\oplus {\bf C}^n$ are transverse subspaces in ${\bf C}^n\oplus 
{\bf C}^n$.  Now let
${\cal I}$ denote any subspace of ${\bf C}^n\oplus {\bf C}^n$ which is transverse
to ${\cal N}$.  Then we claim that for any $f\in C_0^{\infty}({\cal R}
({\bf C}\backslash {\bf a}))$ there
exists a unique $L^2$ Ising type solution, $F$, to $(m-\dsl)F=f$,
with,
$$(c_0(F),c_0^{*}(F))\in {\cal I}.\eqno (1.26)$$
Existence is simple.  We know that the differential
equation $(m-\dsl)F=f$ has an $L^2$ Ising type solution
with $c_0(F)=0$.  By transversality there is a unique
element $w\in N$ so that,
$$0\oplus c_0^{*}(F)-(c_0(w),c_0^{*}(w))\in {\cal I}.$$
It is clear that $F-w$ is a solution satisfying the 
boundary condition (1.26).  Uniqueness is obvious.

Now we introduce the boundary condition that is 
relevant to the Ising model.  Henceforth, let ${\cal I}$ denote the
subspace of ${\bf C}^n\oplus {\bf C}^n$ which consists of all vectors
of the form $(v,v)$ with $v\in {\bf C}^n$. This corresponds to
the boundary condition,
$$c_0(F)=c_0^{*}(F),\eqno (1.27)$$
which the reader should note is the 
``complexification'' of condition (1.18).
Clearly ${\cal I}$ is $n$ dimensional and the formula (1.6) for
the inner product of $L^2$ solutions shows that any
element $w\in {\cal N}$ which has boundary values in ${\cal I}$ has
an $L^2$ norm which is negative unless the linear
functional $c_0$ vanishes on it. This is enough to show that
the function must be 0 and hence that ${\cal I}\cap {\cal N}=\{0\}$. Thus
${\cal I}$ is transverse to ${\cal N}$ and there is an existence and
uniqueness result for solutions to (1.23) satisfying (1.27).
This is the complexification of the result we really
want.  If $f$ is a real function, that is,
$$f=f^{*}=\left[\matrix{\bar {f}_2\cr
\bar {f}_1\cr}
\right],$$
then it is easy to see that the solution, $F$, satisfying
(1.27) must also be real (i.e., $F^{*}=F$).  In this case
both $c_0(F)$ and $c_0^{*}(F)$ will be real numbers. 

It is straightforward to describe the subtraction $\phi_{\nu}$ that
we make from ${\cal W}_{\nu}$.  Let $\varphi_{\nu}(x)$ denote a real valued $
C^{\infty}$ 
function on
${\bf C}$ which is identically 1 in a small ball centered at $a_{
\nu}$ 
and which vanishes outside a slightly larger ball about
$a_{\nu}$ but still small enough so that $\varphi_{\nu}$ is zero in a 
neighborhood of all the other points $a_{\mu}$ for $\mu\ne\nu$. Define
$$\phi_{\nu}(z)=\varphi_{\nu}(z)\Im w_0(z_{\nu}).$$
Consulting (1.8) we see that ${\cal W}_{\nu}-\phi_{\nu}$ will have local 
expansions at each point $a_{\mu}$ of type (1.18).  Write
$\delta {\cal W}_{\nu}:=m^{{1\over 2}}({\cal W}_{\nu}-\phi_{\nu})
.$ Then using the fact that ${\cal W}_{\nu}$ and $\Im w_0$ 
both satisfy the massive Dirac equation we find,
$$(m-\dsl)\delta {\cal W}_{\nu}=\left[\matrix{0&-2\partial\varphi_{
\nu}\cr
-2\bar{\partial}\varphi_{\nu}&0\cr}
\right]m^{{1\over 2}}\Im w_0(z_{\nu}):=f_{\nu}.\eqno (1.28)$$
The scale factor $m^{{1\over 2}}$ is introduced here so that following
limit exits,
$$\lim_{m\rightarrow 0}m^{{1\over 2}}\Im w_0(z_{\nu})={1\over {\sqrt {
2\pi}i}}\left[\matrix{z_{\nu}^{-{1\over 2}}\cr
-\bar {z}_{\nu}^{-{1\over 2}}\cr}
\right].\eqno (1.29)$$
In terms of $\delta {\cal W}_{\nu}$ the coefficient $mc_1^{\nu}({\cal W}_{
\nu})$ which appears 
in the formula for $d\log\tau_{-}$ is given by,
$$mc_1^{\nu}({\cal W}_{\nu})=\left({{m^{{1\over 2}}}\over {I_{{1\over 
2}}(m\epsilon )}}\right){1\over {2\pi}}\int_0^{2\pi}\left(\delta 
{\cal W}_{\nu}\right)_1(\epsilon e^{i\theta_{\nu}})e^{-i{{\theta_{
\nu}}\over 2}}d\theta_{\nu}.\eqno (1.30)$$
The function $\left(\delta {\cal W}_{\nu}\right)_1$ is the first component of $
\delta {\cal W}_{\nu}$ and 
the integral on the right of (1.30) calculates a (half 
integer) Fourier coefficient of this function on the circle
of radius $\epsilon$ about $a_{\nu}.$  Since 
$$\lim_{m\rightarrow 0}{{m^{{1\over 2}}}\over {I_{{1\over 2}}(m\epsilon 
)}}=\sqrt {{2\over {\epsilon}}}\Gamma\left({3\over 2}\right),$$
it will suffice for our purposes to control the 
$m\rightarrow 0$
convergence of $\delta {\cal W}_{\nu}$ in $L^p(C_{\epsilon}(a_{\nu}
))$ for any $p\ge 1$ 
and all $\nu$. Here $C_{\epsilon}(a_{\nu})$ is the circle of radius $
\epsilon$ about
$a_{\nu}.$ A device that will
be useful for this purpose will be to ``excise'' small 
neighborhoods of the
branch points $a_{\nu}$.  Let $D_{\nu}$ denote a small open disk centered
at $a_{\nu}$ and suppose that the radii of these disks are small
enough so that they are mutually disjoint.  We also 
want $D_{\nu}$ chosen small enough so that the $\varphi_{\nu}$ is 
identically 1 on $D_{\nu}$.  The function $f_{\nu}$ will then have its 
support outside the union of the $D_{\nu}$.
Write 
$C_{\nu}=\partial D_{\nu}$ for the {\it clockwise\/} oriented circle that is
the boundary of $D_{\nu}.$  Let $D_{\infty}$ denote a disk centered at
$a_0$ with a radius that is large enough so that all the 
disks $D_{\nu}$ are contained in the interior of $D_{\infty}$.  We will
write $D_{\infty}^c$ for the complement in ${\bf C}$ of $D_{\infty}$.  Write 
$C_{\infty}$ for the {\it counterclockwise\/} oriented circle $\partial 
D_{\infty}.$ 
Let ${\cal D}$ denote the bounded domain which is the 
complement of the union of the closed disks $\bar {D}_{\nu}$ for $
\nu =1,\ldots n$
in $D_{\infty}$.  We now rephrase the problem we need to solve 
to find $\delta {\cal W}_{\nu}$.  We wish to find a solution to
$$(m-\dsl)\delta {\cal W}_{\nu}(x)=f_{\nu}(x),$$
for $x\in {\cal D}$ with the restriction of $\delta {\cal W}_{\nu}$ to $
C_{\mu}$ for 
$\mu =1,\ldots ,n$ equal to the boundary value of a solution
to the Dirac equation in $D_{\mu}$ with local expansion of
type (1.18) (we will refer to this subspace of boundary
values as $W_{\mu}^{(m)}$).  We also require that the restriction of
$\delta {\cal W}_{\nu}$ to $C_{\infty}$ belongs to the space of boundary values 
of solutions to the Dirac equation which are in
$L^2(D_{\infty}^c)$ (we refer to this subspace as $W_{\infty}^{(m
)}$). Of course,
we really need to work in a two fold covering space
for ${\cal D}$ and two fold covers of the circles $C_{\alpha}$, but for
simplicity of exposition in the remainder of this note
we will ignore this (the details of the argument we only 
sketch here will appear in a paper that is still in 
preparation).  Our principal tool in understanding
the $m\rightarrow 0$ limit of the solution to this problem is a 
``guess'' for the $m=0$ Green function.  Here is the guess,
$$G_0(x,y)=-{1\over {4\pi i}}\left[\matrix{\sum_{\nu}u_{\nu}(x)\overline {
v_{\nu}(y)}&g(x,y)\cr
\overline {g(x,y)}&\sum_{\nu}\overline {u_{\nu}(x)}v_{\nu}(y)\cr}
\right],$$
where
$$u_{\nu}(x)=(x-a_{\nu})^{-{1\over 2}}\prod_{\mu\ne\nu}{{(x-a_{\mu}
)^{{1\over 2}}}\over {(a_{\nu}-a_{\mu})^{{1\over 2}}}},$$
$$g(x,y)=\sum_{|\epsilon |=0}c(\epsilon ){{\prod_{\nu}(x-a_{\nu})^{
\epsilon_{\nu}}(y-a_{\nu})^{-\epsilon_{\nu}}}\over {y-x}},$$
with $\epsilon =(\epsilon_1,\epsilon_2,\ldots ,\epsilon_n)$ and each $
\epsilon_{\nu}=\pm{1\over 2}$.  Also
$$|\epsilon |:=\sum_{\nu}\epsilon_{\nu},$$

$$c(\epsilon ):={{\prod_{\mu <\nu}|a_{\mu}-a_{\nu}|^{2\epsilon_{\mu}
\epsilon_{\nu}}}\over {\sum_{|\epsilon |=0}\prod_{\mu <\nu}|a_{\mu}
-a_{\nu}|^{2\epsilon_{\mu}\epsilon_{\nu}}}},$$
and 
$$v_{\nu}(x)=(x-a_{\nu})^{-{1\over 2}}\sum_{|\epsilon (\nu )|=0,\epsilon_{
\nu}={1\over 2}}c(\epsilon )\prod_{\mu\ne\nu}{{(x-a_{\mu})^{\epsilon_{
\mu}}}\over {(a_{\nu}-a_{\mu})^{\epsilon_{\mu}}}}.$$

A possible surprise in this guess are the ``chiral 
symmetry breaking terms'' on the diagonal in $G_0$.  
The $m=0$ Dirac operator $\dsl$ is purely off diagonal
and one might expect the same for its Green function.
However, the boundary conditions we are interested in
force the off diagonal terms even in the $m=0$ limit.
Indeed the principal ingredient that went into this guess
was setting the boundary conditions for the $m=0$ limit
equal to the limits, $W_{\nu}^{(0)}$, of the subspaces $W_{\nu}^{
(m)}$ as 
$m\rightarrow 0$.  It is not difficult to check that $G_0$ is a Green 
function for $-\dsl$ on ${\cal D}$ essentially using the fact that
${1\over {x-y}}$ is a Green function for $\bar{\partial}$.
As a first approximation to $\delta {\cal W}_{\nu}$ we invert $m-\dsl$ 
according to the following scheme,
$$(m-\dsl)^{-1}=-\dsl^{-1}(1-m\dsl^{-1})^{-1}=G_0(1+mG_0)^{-1}.\eqno 
(1.31)$$
It is a classical result [5] that the Cauchy kernel ${1\over {x-y}}$ 
defines a bounded operator on $L^p({\cal D})$ for all $p>2$.  This
result carries over without difficulty to $G_0(x,y)$ (modulo
the difference that we should be working on a covering of ${\cal D}$).
The Neuman series for $(1+mG_0)^{-1}$ then converges in the
strong operator topology on $L^p({\cal D})$ and we can use this 
to show that (1.31) does invert $m-\dsl$ for all small 
enough $m$.  Our first approximation to $\delta {\cal W}_{\nu}$ we take to be
$$\delta_1{\cal W}_{\nu}:=G_0(1+mG_0)^{-1}f_{\nu}.$$
This solves the appropriate inhomogeneous Dirac equation 
but the boundary values are in the subspaces $W_{\nu}^{(0)}$ 
instead of $W_{\nu}^{(m)}$.  The $m\rightarrow 0$ limit of $\delta_
1{\cal W}_{\nu}$ is 
straightforward to compute.

The Calderon projector [2] associated with
the Green function $G_0(x,y)$ maps the direct sum $\oplus W_{\nu}^{
(m)}$
onto $\oplus W_{\nu}^{(0)}$ along the subspace, {\bf N}, of boundary values of
solutions to $\dsl\psi =0$.  For small $m$ one can show that
this projection determines an isomorphism of $\oplus W_{\nu}^{(m)}$ with
$\oplus W_{\nu}^{(0)}$ with a complementary projection on {\bf N} which is $
O(m)$.
The fact that this is true for the Green function $G_0$ is
precisely what makes our choice of $G_0$ suitable in this
application.
Thus we can adjust $\delta_1{\cal W}_{\nu}$ by an element $\delta_
2{\cal W}_{\nu}\in {\bf N}$ which
is $O(m)$ so that $\delta_1{\cal W}_{\nu}+\delta_2{\cal W}_{\nu}$ has boundary values in
$W_{\nu}^{(m)}$.  This is not quite the solution we want but
since $\delta_2{\cal W}_{\nu}\in {\bf N}$ it follows that
$$(m-\dsl)\left(\delta_1{\cal W}_{\nu}+\delta_2{\cal W}_{\nu}\right
)=f_{\nu}+m\delta_2{\cal W}_{\nu}.$$
Thus the adjustment we need to get $\delta {\cal W}_{\nu}$ from
$\delta_1{\cal W}_{\nu}+\delta_2{\cal W}_{\nu}$ is given by the solution to the Dirac
equation with right hand side $m\delta_2{\cal W}_{\nu}=O(m^2)$ and
boundary values in $\oplus W_{\nu}^{(m)}.$ We don't have very
detailed control of the solution to this problem but
because the right hand side is $O(m^2)$ a simple apriori
estimate shows that this last adjustment is $O(m)$ as
well.  Making use of the explicit form for the 
$m\rightarrow 0$ limit of $\delta_1{\cal W}_{\nu}$ we are now in a position to calculate the
limiting form of the logarithmic derivative of $\tau_{-}$.
We find,
$$\lim_{m\rightarrow 0}d\log\tau_{-}={1\over 2}d\log\sum_{|\epsilon 
|=0}\prod_{\mu <\nu}|a_{\mu}-a_{\nu}|^{2\epsilon_{\mu}\epsilon_{\nu}}
.\eqno (1.32)$$
This is the Luther--Peschel formula for the $2n$ point
function.
\vskip.2in\par\noindent
\centerline{\bf References}
\vskip.1in\par\noindent
[1] Anosov, D.V., Bolibruch, A.A., {\sl The Riemann-Hilbert
problem, Aspects of Mathematics}: E; Vol. 22 (1994).
\vskip.1in\par\noindent
[2] Booss--Bavnbek, B., Wojciechowski, K. {\sl Elliptic 
boundary problems for Dirac operators}, Birkh\"auser, 
Boston (1993)
\vskip.1in\par\noindent
[3] Deift, P., Its, A., Zhou, X., {\sl A Riemann-Hilbert approach to 
asymptotic problems arising in the theory of random matrix models,
and also in the theory of integrable statistical mechanics }, Annals of 
Math., {\bf 146}, (1997), pp.149-237.
\vskip.1in\par\noindent
[4] Di Francesco, P., Mathieu, P., S\'en\'echal, D., 
{\sl Conformal Field Theory} , Springer Verlag, New York, (1997)
IBSN 0-387-94785-X
\vskip.1in\par\noindent
[5] Imayoshi, Y., and Taniguchi, M. {\sl An introduction to 
Teichm\"uller spaces}, Springer Verlag, Tokyo, (1992)
ISBN 0-387-70088-9
\vskip.1in\par\noindent
[6] Luther, A., and Peschel, I., {\sl Calculation of critical 
exponents in two dimensions from quantum field theory in
one dimension }, Phys. Rev. B, {\bf 12}, (1975) p. 3908--3917.
\vskip.1in\par\noindent
[7] Palmer, J., and Tracy, C., {\sl Two dimensional Ising correlations: 
convergence of the scaling limit}, Adv. Appl. Math., {\bf 2}, 
(1981) p. 329
\vskip.1in\par\noindent
[8] Palmer, J., and Tracy, C., {\sl Two dimensional Ising correlations: the 
SMJ analysis }, Adv. Appl. Math., {\bf 4}, (1983) p.46--102.
\vskip.1in\par\noindent
[9] Palmer, J., {\sl Tau functions for the Dirac operator in 
the Euclidean plane }, Pacific Journal of Mathematics, 
{\bf 160}, No. 2, (1993) p. 259-342.
\vskip.1in\par\noindent
[10] Palmer, J., Beatty, M., Tracy, C., {\sl Tau functions for 
the Dirac operator in the Poincar\'e disk }, 
Communications in Mathematical Physics, {\bf 165}, No. 1, 
(1994) p. 97-173.
\vskip.1in\par\noindent
[11] Sato, M., Miwa, T., Jimbo, M., {\sl Holonomic quantum 
fields I-V }, Publ. RIMS Kyoto Un., {\bf 14}, (1978) p. 223--267, 
{\bf 15}, (1979) p. 201--278, {\bf 15}, (1979) p. 577--629, {\bf 15}, (1979) p. 
871--972, {\bf 16}, p. 531--584. 
\vskip.1in\par\noindent
[12] Tracy, C., {\sl Asymptotics of a tau function arising in 
the two dimensional Ising model }, Communications in 
Mathematical Physics, {\bf 142}, (1991) p. 297--311.
\vskip.1in\par\noindent
[13] Tracy, C., and Widom, H. {\sl Asymptotics of a class of 
solutions to the cylindrical Toda equations }, 
Communications in Mathematical Physics, {\bf 190}, (1998) p. 
697--721.
\vskip.1in\par\noindent
[14] Wu, T.T., McCoy, B., Tracy, C., Barouch, {\sl Spin-spin 
correlation functions for the two dimensional Ising model: 
Exact theory in the scaling region }, Phys. Rev. B, {\bf 13}, 
(1976) p. 316-374.
\vskip.1in\par\noindent
[15] Wu, T.T. ,{\sl Theory of Toeplitz determinants and the spin correlations
of the two-dimensional Ising model I}, Phys. Rev. {\bf 149 }
(1966), 380-401.
\bye